==================

\documentstyle[12pt]{article}

\begin{document}

\begin{titlepage}
\title{The Hidden Quantum Group of the 8--vertex Free Fermion
Model: q--Clifford Algebras}

\author{R. Cuerno, C. G\'omez, E. L\'opez--Manzanares and G. Sierra \\
{\it Instituto de Matem\'aticas y F\'{\i}sica Fundamental, CSIC} \\
{\it Serrano 123, E--28006 Madrid, SPAIN}}

\date{}

\maketitle

\begin{abstract}
We prove in this paper that the elliptic $R$--matrix of the
eight vertex free fermion model is the intertwiner
$R$--matrix of a quantum deformed Clifford--Hopf algebra. This
algebra is constructed by affinization of a quantum Hopf
deformation of the Clifford algebra.
\end{abstract}

\vskip-12.0cm
\rightline{{\bf IMAFF-2/93}}
\rightline{{\bf February 1993}}
\vskip2cm

\end{titlepage}

\section{Introduction}

The realm of two dimensional integrable models contains two
important families associated to the six vertex and eight vertex
solutions to the Yang-Baxter equation \cite{Bax}.

Whereas the family of six vertex solutions (six vertex model and
their higher spin descendants) are $R-$matrix intertwiners for
different finite dimensional irreducible representations of
$U_q(\widehat{Sl(2)})$, the elliptic eight vertex solutions do not
admit, for the time being, the interpretation as quantum group
intertwiners. To find such a quantum group interpretation of the
eight vertex model would provide, for instance, a natural way to
extend the known hidden quantum group structure of conformal
field theories \cite{CG} to $q-$conformal field theories defined by the
$q-$deformed Knizhnik-Zamolodchikov equation \cite{FR}.

A special class of solutions to the vertex Yang-Baxter (YB) equation
are the ones satisfying the so called free fermion condition:
\begin{equation}
R_{00}^{00}(u)R_{11}^{11}(u)+R_{01}^{10}(u)R_{10}^{01}(u)=
R_{00}^{11}(u)R_{11}^{00}(u)+R_{01}^{01}(u)R_{10}^{10}(u)
\label{2}
\end{equation}

In the six vertex case, $R_{00}^{11}(u)\!=\!R_{11}^{00}(u)\!=\!0$
the solutions to YB satisfy (\ref{2}) and are given by the $R-$matrix
intertwiners of the Hopf subalgebra $U_{\hat q}(\widehat{gl(1,1)})$
\cite{SR}. These intertwiners can be mapped into the ones
of $U_q(\widehat{sl(2)})$ ($q^4 \!=\!1$) for non classical nilpotent
irreducible representations \cite{ACG} with $\hat{q} \!=\! \lambda$
and $\lambda^2$ the eigenvalue of the casimir $K^2$. The
physical interest of the free fermion six vertex solutions is
their close connection with $N\!=\!2$ integrable models. In fact we
can define using the generators of $U_q(\widehat{sl(2)})$ for $q^4 \!=\!1$
a $N\!=\!2$ supersymmetric algebra \cite{B}, and in this case
the free fermion condition
(\ref{2}) reflects the $N\!=\!2$ invariance of the $R-$matrix.
Moreover  the $N\!=\!2$ piece of the solitonic $S-$matrix for the $N\!=\!2$
Ginzburg-Landau superpotential $W\!=\!X^{N+1}/(N+1)\!-\! \beta X$
\cite{FI} can be shown to be given by the intertwiners of
$U_{\hat{q}}(\widehat{gl(1,1)})$ with $\hat{q}^N \!=\!1$.

In the eight vertex case, $R_{00}^{11}(u) \! \neq \!0$
$R_{11}^{00}(u) \! \neq \!0$, solutions to YB satisfying (\ref{2}) have
been known for a long time \cite{FW}. The most general solution
corresponding to imposing non-zero field \cite{F}, \cite{BS}
depends on three spectral parameters $u$, $\psi_1$, $\psi_2$,
and is given by:
\begin{eqnarray}
a&\equiv &R_{00}^{00}=1-e(u)e(\psi_1)e(\psi_2) \nonumber \\
\tilde{a}&\equiv &R_{11}^{11}=e(u)-e(\psi_1)e(\psi_2) \nonumber \\
b&\equiv &R_{01}^{10}(u)=e(\psi_1)-e(u)e(\psi_2) \label{3} \\
\tilde{b}&\equiv &R_{10}^{01}=e(\psi_2)-e(u)e(\psi_1) \nonumber \\
c&\equiv &R_{01}^{01}=R_{10}^{10}=(e(\psi_1)sn(\psi_1))^{1/2}
(e(\psi_2)sn(\psi_2))^{1/2}(1-e(u))/sn(u/2) \nonumber \\
d&\equiv &R_{00}^{11}=R_{11}^{00}=-ik(e(\psi_1)sn(\psi_1))^{1/2}
(e(\psi_2)sn(\psi_2))^{1/2}(1+e(u))sn(u/2) \nonumber
\end{eqnarray}

\noindent
with $e(u)$ the elliptic exponential:
\begin{equation}
e(u)=cn(u)+isn(u)
\label{4}
\end{equation}

\noindent
and $k$ the elliptic modulus. The Yang--Baxter equation
satisfied by this $R$ matrix is \cite{BS}:
\begin{eqnarray}
& & ({\bf 1} \otimes R(u;\psi_1,\psi_2))(R(u+v;\psi_1,\psi_3)
\otimes {\bf 1})({\bf 1} \otimes R(v;\psi_2,\psi_3)) = \nonumber \\
& & (R(v;\psi_2,\psi_3) \otimes {\bf 1})({\bf 1} \otimes
R(u+v;\psi_1,\psi_3))(R(u;\psi_1,\psi_2) \otimes {\bf 1})
\label{1}
\end{eqnarray}

\noindent
The simplest way
to catch the physical meaning of solution (\ref{3}) is to define
the corresponding spin chain hamiltonian:
\begin{equation}
\left. H= \sum_{j=1}^{N}i \frac{\partial}{\partial u}R_{j,j+1}(u;\psi,\psi)
\right|_{u=0}
\label{5}
\end{equation}

\noindent
which is the well known $XY-$ model in an external magnetic
field \cite{LSM}:
\begin{equation}
H= \sum_{j=1}^{N} [ (1+ \Gamma) \sigma_j^x \sigma_{j+1}^x +
(1- \Gamma) \sigma_j^y \sigma_{j+1}^y + h(\sigma_j^z +
\sigma_{j+1}^z) ]
\label{6}
\end{equation}

\noindent
where:
\begin{eqnarray}
\Gamma & = & \frac{2cd}{ab+ \tilde{a} \tilde{b}} = ksn(\psi)
\label{7} \\
h & = & \frac{a^2+b^2- \tilde{a}^2- \tilde{b}^2}{2(ab+\tilde{a}
\tilde{b})}=cn(\psi) \nonumber
\end{eqnarray}

In this letter and as a preliminary step of the long term
process of finding the quantum group symmetry of the eight vertex
model, we will define a fully fledged Hopf algebra such that
its $R-$intertwiners coincide with the elliptic free fermionic
eight vertex solution (\ref{3}).

\section{The quantum Clifford algebra}

A Clifford algebra $C(\eta)$ associated to a cuadratic form
$\eta$ is the associative algebra generated by the elements $\{
\Gamma_{\mu} \}_{\mu =0}^D$, which satisfy:
\begin{equation}
\{ \Gamma_{\mu}, \Gamma_{\nu} \} = 2 \eta_{\mu \nu} {\bf 1}
\;\;\; \mu = 1,\ldots,D
\label{8}
\end{equation}

Associated to $C(\eta)$ we define the Clifford--Hopf algebra
CH(D) as the associative algebra generated by $\Gamma_{\mu}
\;\;\;(\mu=1,\ldots,D),\;\; \Gamma_{D+1}$ and the central elements
$E_{\mu} \;\;\; (\mu=1,\ldots,D)$ satisfying the following relations:
\begin{eqnarray}
& & \Gamma_{\mu}^2 = E_{\mu} \;\; , \;\; \Gamma_{D+1}^2 = {\bf 1}
\nonumber \\
& & \{ \Gamma_{\mu}, \Gamma_{\nu} \} =0, \;\; \mu \neq \nu
\nonumber \\
& & \{\Gamma_{\mu}, \Gamma_{D+1} \} =0 \label{9} \\
& & [ E_{\mu}, \Gamma_{\nu} ] = [ E_{\mu}, \Gamma_{D+1} ] =
[ E_{\mu}, E_{\nu} ] = 0 \;\; \forall \mu, \nu \nonumber
\end{eqnarray}

\noindent
The algebra CH(D) is a Hopf algebra with the following
comultiplication $\Delta$, antipode $S$ and counit $\epsilon$:
\begin{equation}
\begin{array}{lll}
\Delta (E_{\mu}) = E_{\mu} \otimes {\bf 1} + {\bf 1} \otimes
E_{\mu}, & S(E_{\mu}) = -E_{\mu}, &\epsilon(E_{\mu}) = 0 \\
\Delta (\Gamma_{\mu}) = \Gamma_{\mu} \otimes {\bf 1} + \Gamma_{D+1}
 \otimes \Gamma_{\mu}, & S(\Gamma_{\mu}) = \Gamma_{\mu}
\Gamma_{D+1}, & \epsilon(\Gamma_{\mu}) = 0 \\
\Delta(\Gamma_{D+1}) = \Gamma_{D+1} \otimes \Gamma_{D+1}, &
S(\Gamma_{D+1}) = \Gamma_{D+1}, & \epsilon(\Gamma_{D+1}) = 1 \\
\end{array}
\label{10}
\end{equation}

\noindent
For D even the elements $E_{\mu} \;\;\; (\mu=1,\ldots,D)$
and the product $\Gamma_1
\cdots \Gamma_D \Gamma_{D+1}$ are casimirs of CH(D), therefore
in an irreducible representation of CH(D) we get $E_{\mu} =
\eta_{\mu \mu}$, and $\Gamma_{D+1} \sim \Gamma_1 \cdots
\Gamma_D$ which means that the irreps of CH(D) are isomorphic to
those of $C(\eta)$ for all possible signatures of $\eta$ (there
is a unique faithful representation of $C(\eta)$ of dimension $2^D$). For D odd
similar arguments show that the representation theory of CH(D)
is related to that of $C(\eta)$ for $\eta$ a quadratic form
defined in one dimension higher, namely D+1.

The quantum deformation of CH(D), that we will denote
CH$_q$(D), is defined by:
\begin{equation}
\Gamma_{\mu}^2 = \left[ E_{\mu} \right]_q =
\frac{q^{E_{\mu}}-q^{-E_{\mu}}}{q- q^{-1}}
\label{11}
\end{equation}

\noindent
with the rest of equations (\ref{9}) unchanged. The
comultiplication for $\Gamma_{\mu}$ is now given by:
\begin{equation}
\Delta \Gamma_{\mu} = \Gamma_{\mu} \otimes q^{-E_{\mu}/2} +
q^{E_{\mu}/2} \Gamma_{D+1} \otimes \Gamma_{\mu}
\label{12}
\end{equation}

\noindent
The Hopf algebra CH$_q$(D) for $D=2$ is very close to the two
parameter quantum supergroup ${\cal U}_{\alpha, \beta}(su(1,1))$
defined in \cite{R}. The correspondence between both algebras
is given by the substitutions $ q^{E_x} \longrightarrow
\alpha^{E}$ and $ q^{E_y} \longrightarrow
\beta^{E}$. Notice that in $CH_q(2)$ we have two central
elements $E_x, E_y$ and one quantum deformation parameter,
while in ${\cal U}_{\alpha, \beta}(su(1,1))$ there exist one
central element and two parameters. This difference will be
important in the representation theory. We observe that the
``SUSY grading" is played in our case by $\Gamma_3$, and in general
by $\Gamma_{D+1}$ for $D>2$.

Next we proceed to define a sort of affinization of the Hopf
algebra CH$_q$(D). The generators of this new algebra that we
denote $\widehat{CH_q(D)}$ are: $E_{\mu}^{(i)}, \Gamma_{\mu}^{(i)},
\Gamma_{D+1}^{(i)}, \;\;i=0,1$ satisfying (\ref{11}) and (\ref{12})
for each value of $i$.
In what follows we will consider only the case $D=2$\footnote{To define
the algebra $\widehat{CH_q(D)}$ properly we should add to (\ref{11}) the
equivalent to Serre's relations. These will not be relevant for
the discussion in this paper.}.

A two dimensional irrep $\pi_{\xi}$ of $\widehat{CH_q(D)}$
is labelled by three complex parameters $\xi = (z,\lambda_x,\lambda_y) \in
C_{\times}^3$ and reads:
\begin{eqnarray}
\pi_{\xi}(\Gamma^{(0)}_x) =
\left(\frac{\lambda_x^{-1}-\lambda_x}{q-q^{-1}}\right)^{1/2} \left(
\begin{array}{cc} 0 & z^{-1} \\ z & 0 \\ \end{array} \right) &, &
\pi_{\xi}(\Gamma^{(1)}_x) =
\left(\frac{\lambda_x-\lambda_x^{-1}}{q-q^{-1}}\right)^{1/2} \left(
\begin{array}{cc} 0 & z \\ z^{-1} & 0 \\ \end{array} \right)
\nonumber \\
& & \nonumber \\
\pi_{\xi}(\Gamma^{(0)}_y) =
\left(\frac{\lambda_y^{-1}-\lambda_y}{q-q^{-1}}\right)^{1/2} \left(
\begin{array}{cc} 0 & -i z^{-1} \\ i z & 0 \\ \end{array} \right) &, &
\pi_{\xi}(\Gamma^{(1)}_y) =
\left(\frac{\lambda_y-\lambda_y^{-1}}{q-q^{-1}} \right)^{1/2}\left(
\begin{array}{cc} 0 & -i z \\ i z^{-1} & 0 \\ \end{array}
\right) \nonumber \\
& & \nonumber \\
\pi_{\xi}(\Gamma^{(0)}_3) =
\left( \begin{array}{cc} 1 & 0 \\ 0 & -1 \\ \end{array} \right)
&, & \pi_{\xi}(\Gamma^{(1)}_3) =
\left( \begin{array}{cc} 1 & 0 \\ 0 & -1 \\ \end{array} \right)
\label{12'} \\
& & \nonumber \\
\pi_{\xi}(q^{E^{(0)}_x}) = \lambda_x^{-1} & , &
\pi_{\xi}(q^{E^{(1)}_x}) = \lambda_x \nonumber \\
& & \nonumber \\
\pi_{\xi}(q^{E^{(0)}_y}) = \lambda_y^{-1} &, & \pi_{\xi}(q^{E^{(1)}_y}) =
\lambda_y \nonumber
\end{eqnarray}

\noindent
The intertwiner $R_{\xi_1,\xi_2}$ for two of these irreps is
defined by the condition
\begin{equation}
R_{\xi_1 \xi_2} \Delta_{\xi_1 \xi_2}(a) = \Delta_{\xi_2
\xi_1}(a) R_{\xi_1 \xi_2} \;\;\; \forall a \in \widehat{CH_q(2)}
\label{14}
\end{equation}

\noindent
with $\Delta_{\xi_1 \xi_2}= \pi_{\xi_1} \otimes \pi_{\xi_2}
(\Delta)$. Assuming that $R_{\xi_1,\xi_2}$ is an invertible
matrix, then the intertwiner equation (\ref{14}) implies:
\begin{equation}
{\rm tr} \;\Delta_{\xi_1 \xi_2}(a) = {\rm tr} \;\Delta_{\xi_2
\xi_1}(a) \;\;\; \forall a \in \widehat{CH_q(2)}
\label{15}
\end{equation}

\noindent
For $a= \Gamma_x^{(0)} \Gamma_y^{(1)}$ we obtain the following
constraint on the labels of the irreps which admit an intertwiner:
\begin{equation}
\frac{2 (\lambda - \mu)}{(1-\lambda^2)^{1/2}
(1-\mu^2)^{1/2}(z^2 - z^{-2})} = k
\label{16}
\end{equation}

\noindent
with $k$ an arbitrary constant. Equation (\ref{16}) defines a
two dimensional variety embedded in ${\bf C}^{3}$ which can be
uniformized in terms of elliptic functions. Identifying $k$ in
(\ref{16}) with the elliptic modulus we define a new variable
$\varphi$ by:
\begin{equation}
z^2 = cn(\varphi) + i sn(\varphi)
\label{17}
\end{equation}

\noindent
Definig now:
\begin{equation}
\lambda_x = \tanh x \;\; , \;\; \lambda_y = \tanh y
\label{18}
\end{equation}

\noindent
equation (\ref{16}) becomes
\begin{equation}
e^{x-y} = dn(\varphi) + i k sn(\varphi)
\label{19}
\end{equation}

\noindent
which means that $x+y$ is independent of $\varphi$, therefore
each point in the curve (\ref{16}) can be parametrized by two
complex parameters $(\varphi, \psi)$ with $\psi$ defined by
\begin{equation}
\tanh \left( \frac{x+y}{2} \right) = cn(\psi) + i sn(\psi)
\label{20}
\end{equation}

\noindent
The main result of this paper is that, given two irreps lying on
the same curve (\ref{16}),
$\xi_1\;(\varphi_1, \psi_1)$ and $\xi_2\;(\varphi_2, \psi_2)$ the
intertwiner $R$--matrix $R_{\xi_1, \xi_2}$ coincides with the one given in
(\ref{3}) (up to a diagonal change of basis)
provided we identify $u=\varphi_1-\varphi_2$. Notice
from (\ref{17}) that the ``affine" parameter $z$ becomes the
standard exponential in the trigonometric limit. The derivation
of (\ref{3}) is long but straighforward, and we have used the following
identity
among elliptic functions:
\begin{equation}
e(\varphi_1-\varphi_2) = \frac{e(\varphi_1)
(dn(\varphi_1)+1) (dn(\varphi_2)+1) -k^2 e(\varphi_2) sn(\varphi_1)
sn(\varphi_2)}{e(\varphi_2) (dn(\varphi_1)+1) (dn(\varphi_2)+1)
-k^2 e(\varphi_1) sn(\varphi_1) sn(\varphi_2)}
\label{21}
\end{equation}

Summarizing our results we have proved that the intertwiner $R$--matrix for
two dimensional irreps of the Hopf algebra $\widehat{CH_q(2)}$
is the free fermion eight vertex solution to the Yang--Baxter equation.

\section{Comments}

The Sklyanin algebra \cite{S} of the eight vertex model is
determined by the corresponding elliptic curve and the
anisotropy $\gamma$ \cite{Ch}. In the free fermionic case, i.e.
$\gamma = K$, the curve, for the most general case with
non--zero field, is given by (\ref{16}). An important question
that we will address in a future publication, is the
mathematical meaning, inside $\widehat{CH_q(2)}$, of the
$\gamma=K$--Sklyanin algebra.

Taking into account that the trigonometric limit of the free
fermion model is given by the six vertex free fermion model and
that this $R-$matrix is the intertwiner of
$U_{\hat q}(\widehat{gl(1,1)})$, it is plausible to conjecture
that the Hopf algebra $\widehat{CH_q(2)}$ plays the role, in the
sense of reference \cite{FR}, of hidden quantum group of the
$q-$WZW model defined by $U_q(\widehat{gl(1,1)})$. More
precisely we expect that the connection matrices of the $q-$KZ
equation for $U_q(\widehat{gl(1,1)})$ are quantum $6-j$ symbols
of $\widehat{CH_q(2)}$.

Another interesting issue is the interpretation of the eight
vertex free fermion $R-$matrix as an scattering $S-$matrix in
the sense of Zamolodchikov \cite{Z}. From our previous results
we know that the correspondent ``solitons" define now irreducible
representations of $\widehat{CH_q(2)}$. Even though we cannot
expect a field theory limit preserving the elliptic nature of
this $S-$matrix, as a consequence of the c-theorem \cite{ZDV},
the elliptic $S-$matrix (\ref{3}) may still have a good physical
meaning in the lattice, maybe related to the dynamics of the
cnoidal waves in a Toda lattice \cite{T}.

Finally and based on the previously mentioned close connection
between $N\!=\!2$ soliton $S-$matrices and intertwiners of
$U_q(\widehat{gl(1,1)})$, it is natural to wonder if some
relevant information on $N\!=\!2$ integrable models is still
hidden in the quantum Clifford algebra $\widehat{CH_q(2)}$.

\subsection*{Acknowledgements}

We would like A. Berkovich for valuable discussions. The work of
R.C. and E.L.M. is supported by M.E.C. fellowships.

\end{document}